\documentstyle[epsfig,psfig]{aipproc}

\def\to{\rightarrow}
\def\mpl{\ifmmode \overline M_{Pl}\else $\bar M_{Pl}$\fi}
\def\epem{\ifmmode e^+e^-\else $e^+e^-$\fi}

\begin{document} 

\title{New Physics Working Group Summary of LCWS2000@FNAL} 
 
\author{C.~S.~ Kim
\footnote{kim@kimcs.yonsei.ac.kr, ~ cskim@pheno.physics.wisc.edu}
}

\address{Department of Physics and IPAP, Yonsei University, Seoul 120-749,
Korea}   
\maketitle

\begin{abstract} 
Here I summarize P5-WG (New/Alternative Physics Working Group) of
LCWS2000@FNAL,  held on Oct 24-28, 2000.
There were 13 talks altogether, 7 talks on collider signals from new
particles/interactions and 6 talks on extra dimensional physics.
We had very active and hot discussions among participants for those
new/alternative physics/ideas. 
In bottom table, I show those 13 speakers'  names and titles.
\end{abstract} 
 
\section*{I. Introduction} 

\begin{table}[tb] 
%\begin{table}[htbp] 
\caption{Who are the speakers with new ideas? } 
%\centerline{\footnotesize\smalllineskip 
\begin{tabular}{ll} 
%\hline 
Collider Speakers & Titles \\ \hline 
D. Dominici & Signals of new vector resonances \cite{Dom}  \\
P. Kalyniak & Discovery and identification of $W'$ \cite{Kal} \\
S. Riemann & New physics in fermion pair production \cite{Rie} \\
M. Battaglia & Direct search of $Z'$ at CLIC \cite{Bat} \\
V.A. Ilyin & Potentials of LC in stoponium searches \cite{Ily} \\
T. Han & Higgs-gauge boson couplings with CP violation \cite{Han} \\
C. Heusch & Finding heavy Majorana neutrinos in LC \cite{Heu} \\  \hline\hline
ED Speakers & Titles \\ \hline
J. Lykken & Search for Maximal Weirdness \cite{Lyk} \\
J. Heweet & Signals of non-commutative field theories \cite{Hew} \\
H. Davoudiasl & Probing geometry of the universe at NLC \cite{Dav} \\ 
T. Rizzo & Probing RS Warped ED signals at LC \cite{Riz} \\ 
T. Takeuchi & Universal torsion induced interactions from ED \cite{Tak} \\
H.C. Cheng & Electroweak symmetry breaking and ED \cite{Che} \\ \hline
\end{tabular} 
%\end{tabular}} 
\end{table} 

On the first day of the Workshop, 
Komamiya asked, in his talk on ``Chages to Participants",
to our P5-WG (New/Alternative Physics Working Group)
to investigate New and Crazy Ideas. How crazy is really crazy? Who are those 13
speakers with new ideas? Please look at next table for the speakers and
titles. 
\\

Followings
are the list of really remarkable ideas: 
\begin{itemize}
\item Planck scale on your finger tip? 
\item Natural composite Higgs and EWSB? 
\item Dynamical symmetry breaking and BESS? 
\item $W',~Z'$, heavy Majorana neutrinos?
\item Geometry of the universe and universal torsion?
\item CP-violationg Higgs-gauge boson couplings?
\item non-commutative field theory?
\item Maximal weirdness?
\end{itemize}
As you can see easily, the subjects are changing wildly from each
talk,  and the summarizing those all talks would be an almost impossible task. 
In next section, I summarize the talks on signals of
new particles/interactions at future colliders. Then, I summarize the talks
on physics of Extra Dimension (ED). In last section, I comment on possible
violations of CTP and Equivalence Principle.

\section*{II. New Particles/Interactions Signals}

\paragraph*{Study of Degenerate-BESS}\cite{Dom}.
The degenerate BESS model (D-BESS)~\cite{dbess} is a realization of dynamical 
electroweak symmetry breaking with decoupling.  
The D-BESS model introduces two new triplets of gauge bosons, which are 
almost degenerate in mass, ($L^\pm$, $L_3$),
($R^\pm$, $R_3$). The extra parameters  are a new gauge coupling constant
$g''$ and a mass parameter $M$, related to the scale of the
underlying symmetry breaking sector.
In the charged sector the $R^\pm$ fields  are not mixed and $M_{R^\pm}=M$,
while $M_{{L}^\pm}\simeq M (1+x^2)$ where $x=g/g''$ with $g$  the usual 
$SU(2)_W$ gauge coupling constant.
The $L_3$, $R_3$ masses are given by
$M_{L_3}\simeq  M\left(1+ x^2\right),~~ M_{R_3}\simeq
M \left(1+ x^2 \tan^2 \theta\right)$
where $\tan \theta = s_{\theta}/c_{\theta} = g'/g$ and $g'$ is the usual
$U(1)_Y$ gauge coupling constant. These resonances are narrow and almost 
degenerate in mass with $ \Gamma_{L_3}/M\simeq 0.068~ x^2$ and
$\Gamma_{R_3}/M\simeq 0.01~ x^2$, while the neutral mass splitting is:
$\Delta M/M=(M_{L_3}-M_{R_3})/M 
\simeq \left( 1-\tan^2 \theta \right) x^2\simeq 0.70 ~x^2$.
This model respects the existing stringent bounds from electroweak precision 
data since the $S,T,U$ (or $\epsilon_1, \epsilon_2, \epsilon_3$) parameters 
vanish at the leading order due to an additional custodial
symmetry. Therefore, the precision electroweak data only set loose bounds on
the parameter space of the model,
comparable to those from the direct search at the Tevatron~\cite{dbess}.
Future hadron colliders may be able to discover these new
resonances which are produced through a $q \bar q$ annihilation and which 
decay in the leptonic channel $q{\bar q'}\to L^\pm,W^\pm\to (e \nu_e)
\mu\nu_\mu$ and $q{\bar q}\to L_3,R_3,Z,\gamma\to 
(e^+e^-)\mu^+\mu^-$.  
\begin{table}[h]
\caption{Sensitivity to $L_3$ and $R_3$ production at the LHC and CLIC 
for $L=$100(500)~fb$^{-1}$ with $M=$1,2(3)~TeV at LHC and
$L=$1000~fb$^{-1}$ at CLIC.}
\begin{tabular}{c c c c c c c}
$g/g''$ & $M$ & $\Gamma_{L_3}$ & $\Gamma_{R_3}$ & $S/\sqrt{S+B}$&
 $S/\sqrt{S+B}$ & $\Delta M$
\\
& (GeV) &(GeV) & (GeV) & LHC ($e+\mu$) & CLIC (hadrons) &  CLIC \\
 \hline  0.1 &
1000 & 0.7 & 0.1 &17.3 & &
\\
0.2 & 1000 & 2.8 & 0.4 & 44.7& &
\\\hline
0.1 & 2000 & 1.4 & 0.2 &3.7& &  
\\
0.2 & 2000 & 5.6 & 0.8 & 8.8& &
 \\\hline
0.1 & 3000 & 2.0 & 0.3 &(3.4)& ~62 & 23.20 $\pm$ .06
\\
0.2 & 3000 & 8.2 & 1.2 &(6.6)& 152 & 83.50 $\pm$ .02
 %\\
%\hline
\end{tabular}
\label{dom:table1}
\end{table}  
The relevant observables are the di-lepton transverse and invariant masses.
The main backgrounds, left to these channels after the lepton isolation
cuts, are the Drell-Yan processes with SM gauge bosons
exchange in the electron and muon channel. Results are given in 
Table~\ref{dom:table1} for the combined electron and muon channels
for $L=100$~fb$^{-1}$. Results are given for an integrated luminosity of 
500~fb$^{-1}$ assuming $M=$3~TeV.
The discovery limit at LHC with $L=100$~fb$^{-1}$ is  $M\sim 2$~TeV
with $g/g''=0.1$. Beyond discovery, the possibility to disentangle the double 
peak structure depends strongly on $g/g''$ and smoothly on the mass.
A lower energy LC can also probe this multi-TeV region through 
the virtual effects in the cross-sections for $e^+e^-\to {L_3},{R_3},Z,\gamma
\to f \bar f $. Due to the presence of new spin-one resonances the 
annihilation channel in $f \bar f$ and $W^+W^-$ is more efficient than
the fusion channel. 
In the case of D-BESS, the $L_3$ and $R_3$ states 
are not strongly coupled to $WW$ making the $f\bar f$ final states the most
favorable channel for discovery. 
\\

\paragraph*{Investigation of $W'$ Bosons at LC}\cite{Kal}.
Extra gauge bosons are a fundamental part of many extensions of the 
Standard Model (SM). Extensive investigations of the neutral  
$Z'$ exist in the literature but the charged $W'$ has been less well  
studied. We focus here on the possibility of finding a $W'$ and  
measuring its couplings to fermions at a high energy $e^+e^-$ collider.  
The models (see references in \cite{wp1,wp2}) we consider are the  
Left-Right symmetric model (LRM) based on the gauge group  
$SU(2)_L \times SU(2)_R \times U(1)_{B-L}$, 
the Un-Unified model (UUM) 
based on the gauge group $SU(2)_q \times SU(2)_l  
\times U(1)_Y$ where the quarks and leptons each transform under their own  
$SU(2)$, a Third Family Model (3FM) based on the gauge group $SU(2)_h  
\times SU(2)_l \times U(1)_Y$ where the quarks and leptons of the third  
(heavy) family transform under a separate group \cite{3fm} 
and the KK model (KK) which  
contains the Kaluza-Klein excitations of the SM gauge bosons that 
are a possible consequence of theories with large extra dimensions. We  
also consider a $W'$ with SM couplings as a benchmark which we  
denote as the Sequential Standard Model (SSM).  
The particular processes which we study are 
$e^+e^- \to \nu \bar{\nu} \gamma$ and $e \gamma \to \nu q +X$. 
The first process we consider is $e^+e^- \to \nu \bar{\nu} \gamma$, 
which includes contributions from both $W'$'s and $Z'$'s. The  
kinematic variables of interest are the photon's energy, $E_\gamma$,  
and its angle relative to the incident electron, $\theta_\gamma$, both  
defined in the $e^+e^-$ centre-of-mass frame.  To take into account  
finite detector acceptance we imposed constraints on the kinematic  
variables such that $E_\gamma \geq 10$~GeV and $10^o \leq \theta_\gamma \leq  
170^o$.  The most serious background is radiative Bhabha scattering  
where the scattered $e^+$ and $e^-$ go undetected down the beam pipe.   
We suppress this background by restricting the photon's transverse  
momentum to $p_T^\gamma > \sqrt{s}\sin\theta_\gamma \sin\theta_v  
/(\sin\theta_\gamma +\sin\theta_v )$ where $\theta_v=25$~mrad and is  
the minimum angle to which the veto detectors may observe electrons or  
positrons.  There are also higher order backgrounds which cannot be  
suppressed, such as $e^+e^- \to \nu\bar{\nu} \nu' \bar{\nu}'\gamma$,  
which would have to be included in an analysis of data. 
The second process investigated is $e\gamma\to \nu q+X$, using 
photon spectra from both the Weizsacker Williams process and from a  
backscattered laser. This process contains contributions from only $W'$'s 
and not from $Z'$'s. Starting with the process $e\gamma \to  
\nu q\bar{q}$ the $W'$ contributions can be enhanced  
by imposing the kinematic cut that either the $q$ or $\bar{q}$ is  
collinear to the beam axis.  In this kinematic region the process  
$e\gamma\to \nu q\bar{q}$ is approximated quite well by the simpler  
process $e q\to \nu q'$ where the quark is described by the quark  
parton content of the photon, the so-called resolved photon  
approximation.   
We use the process $eq\to \nu q'$ to obtain our 
limits as it is computationally much faster and the limits obtained in  
this approximation are in good agreement with those using the full  
process. 
\\

\begin{table}[b!]
\caption{Results of the fits for the cross section scan of a $Z'_{SM}$
obtained by assuming no radiation and ISR with the effects of two different
optimization of the CLIC luminosity spectrum.}
\label{table1}
\begin{tabular}{l c c c}
Observable & Breit Wigner & CLIC.01 & CLIC.02 \\ \hline
$M_{Z^{'}}$ (GeV) & 3000 $\pm$ .12  & $\pm$ .15 &  $\pm$ .21 \\
$\Gamma(Z^{'})/\Gamma_{SM}$ & 1. $\pm$ .001 & $\pm$ .003  & $\pm$ .004 \\
$\sigma^{eff}_{peak}$ (fb) & 1493 $\pm$ 2.0 & 564 $\pm$ 1.7 & 669 $\pm$ 2.9 \\
\hline
\end{tabular}
\end{table}

\paragraph*{Study of $Z'$ at CLIC}\cite{Bat}.
The $Z'$ mass and width can be determined by performing either an energy scan,
like the $Z^0$ scan performed at {\sc Lep}/{\sc Slc} and also foreseen
for the $t \bar t$ threshold, or an auto-scan, by tuning the collision energy 
just above the top of the resonance and profiting  of the long tail of the 
luminosity spectrum to probe the resonance peak. For the first method
both di-jet and di-lepton final states can be considered, while for the 
auto-scan only $\mu^+ \mu^-$ final states may provide with the necessary 
accuracy for the $Z'$ energy. $e^+e^- \rightarrow Z'$ events have been 
generated for $M_{Z'}$ = 3~TeV, including the effects of ISR, 
luminosity spectrum and $\gamma \gamma$ backgrounds, assuming SM-like 
couplings, corresponding to a total width $\Gamma_{Z'_{SM}} \simeq$ 90~GeV.
A data set of  1000 fb$^{-1}$ has been assumed for the
CLIC.01 beam parameters and of 400 fb$^{-1}$ for CLIC.02, corresponding to one 
year ( = 10$^{7}$~s) of operation at nominal luminosity. This has been shared
in a 3 to 7 points scan and $M_{Z^{'}}$, $\Gamma(Z^{'})/\Gamma_{SM}$ and 
$\sigma_{peak}$ have been extracted from a $\chi^2$ fit to the predicted
cross section behaviour for different mass and width values. The dilution of 
the analysing power due to the beam energy spread is appreciable, as can be 
seen by comparing the statistical accuracy from a fit to the pure Born cross
section to after including ISR and beamstrahlung effects. Still, the 
relative statistical accuracies are better than 10$^{-4}$ on the mass and 
$5 \times 10^{-3}$ on the width. 
Sources of systematics from the knowledge of the
shape of the luminosity spectrum have also been estimated. In order to keep
$\sigma_{syst} \le \sigma_{stat}$ it is necessary to control 
$N_{\gamma}$ to better than 5\% and the fraction $\cal{F}$ of collisions at 
$\sqrt{s} < 0.995 \sqrt{s_{0}}$ to about 1\%.
\\

\paragraph*{Stoponium Searches at LC}\cite{Ily}.
Search potentials are estimated for stoponium ($S$ - bound 
states of the $t$-quark superpartner), considering the $e^+e^-$ and the 
$\gamma\gamma$ (Photon Linear Collider-PLC) options. The stop bound state can 
be described as a quasistationary system with binding energy $\sim$ 1 GeV 
for $M_S=200-600$~GeV, if the formation process is faster than destroying 
one. The SUSY scenario, where tree-level stop decays are somehow suppressed 
and, therefore, stop decay cannot destroy the stoponium formation, is not 
an exceptional case (see discussions in \cite{Ily,Noj}). 
PLC could be the best machine for discovery of 
these new narrow strong resonances. Thousands of stoponiums can be produced 
per 100 fb$^-1$ in the high energy peak. In the case of scenarios when 
stoponium decays mainly into two gluons the {\it signal/background} ratio 
is about 1/4. In addition the decay channel $S\to hh$ could be seen with 
high significance. Thus, several weeks run is sufficient for the discovery 
of stoponium, if its mass is approximately known (e.g. from observation of 
direct stops production at LHC). Then, in MSSM scenarios with dominant 
$S\to hh$ decay PLC shows excellent possibilities to discover stoponium, 
practically immediately after beginning of operating. 
The signal significance of stoponium events is shown 
in various channels for the MSSM scenarios with $S\to gg$ decay mode being 
dominant, and the total number of stoponium events in case when dominant 
decay channel is $S\to hh$. 
The $e^+e^-$ option also has good discovery prospects but only in the case 
of the second scenario with dominant, with hundreds of events per 100 
$fb^{-1}$. Interesting possibility appears in the case when the resonance 
is seated on 0.1\% width luminosity peak -- one could resolve the stoponium 
exited states.  
\\

\paragraph*{Detection of Heavy Majorana Neutrinos}\cite{Heu}.
As argued for years, a TeV-level $e^-e^-$ collider has the 
unique capability of producing a ``quasi-elastic" back-to-back emitted $W^-$ 
pair after exchanging a $t$-chanel exchanged Majorana neutrino. A careful
investigation of the  relevant parameter space 
led us to come up with respectable counting 
rates if the electrons are both left-handed and exchange a Majorana 
neutrino with mass of order $m_W < \sqrt{s} < m_N$, where $N$ stands for 
the heavy neutrino. The cross section becomes \cite{heuschmink}
\begin{equation} 
\sigma (\ell_L^-\ell_L^- \to W^-W^-);{\cal N} \simeq 
  {1\over M\ {\rm (TeV)}^2} \left( {s\over M^2} \right)^2 
  \left|{U_{ei}^2 \over 4\pi}\right| (4\cdot 10^5\ fb)\ . 
\label{Heu-eq}
\end{equation} 
The mixing parameters $U_ei$, where the index $i$ stands for one of at least 
two heavy neutrinos, are constrained by low-energy rare lepton decays to 
the values $(U_{ei})^2 \times (2-40)10^{-4}$. 
The final-state signatures are 
spectacular, cannot be missed, and are easily distinguished from 
background interactions. Furthermore, any emerging signal must vanish as 
soon as the incoming electron helicity is changed. 
It has been argued by a number of authors that the non-observation 
of neutrinoless double beta decay ought to be interpreted as precluding 
the potential existence of the heavy Majorana neutrinos the Linear 
Collider will search for. This notion rests on the misconception that our 
reaction (\ref{Heu-eq}) is tantamount to ``inverse neutrinoless double beta
decay". In  fact, this nuclear decay, on the quark-lepton level, can happen
only if  two d quarks from two different neutrons inside the parent nucleus 
approach to within $(m_N)^-1$, or some $10^{-16}$cm; it is constrained by the 
color Coulomb hard-core repulsion. A careful estimate of several scenarios 
leads to suppression factors of order $\sim$200, well beyond what would 
preclude a possible observation of the double beta decay in question. 
For detection of Majorana neutrinos at $e-\mu$ colliders, please look
Ref. \cite{emu}.  

%\paragraph*{

%\paragraph*{

\section*{III. Physics with Extra Dimension}

\paragraph*{RS Phenomenolgy at LC}\cite{Dav,Riz}.
Randall and Sundrum(RS){\cite {RS}} have recently proposed a novel approach 
in dealing with the hierarchy problem wherein an exponential warp factor 
arises from a 5-d non-factorizable geometry based on a slice of 
AdS$_5$ space. Here, two 3-branes sit at the orbifold fixed 
points $y=0$ (Plank brane) and $y=\pi r_c$ (SM or TeV brane) with equal 
and opposite 
tensions with the AdS$_5$ space between them. The model contains no large 
parameter hierarchies with $\mpl$, the 5-d Planck scale, $M_5$, and the AdS 
curvature parameter, $k$, being of qualitatively similar magnitudes. TeV 
scales can be generated on the brane at $y=\pi r_c$ if gravity is localized on 
the other brane and $kr_c \simeq 11-12$; indeed in this case the scale of 
physical processes on the SM brane is found to be given by 
$\Lambda_\pi=\mpl e^{-kr_c\pi}$ which is of order a TeV.  
Such a model leads to very interesting and predictive phenomenology that can 
be explored in detail at colliders{\cite {dhr}}. In the simplest scenario the 
SM fields are constrained to lie on the TeV brane while gravitons can 
propagate in the bulk in which case only two 
parameters are necessary to describe the model: $c=k/\mpl$, which is expected 
to be near though somewhat less than unity, and $m_1=kx_1e^{-kr_c\pi}$, 
which is the mass of the first graviton Kaluza-Klein excitation. The masses 
of the higher excitations are given by $m_n=m_1 x_n/x_1$, where the 
$x_n$ are roots of the Bessel function $J_1(x_n)=0$, and are thus not equally 
spaced. While the massless zero mode graviton couples in the usual manner as 
$(\mpl)^{-1}$, the tower states instead couple as $\Lambda_\pi^{-1}$. 
The most distinctive prediction of this scenario is the direct production of 
weak scale graviton resonances at colliders as is shown in Fig.~1 for the case 
of a linear collider. Note that for fixed mass the width of each resonance 
is proportional to $c^2$; for resonances 
beyond the first KK excitation, the width grows as $m^3$. This explains why 
resonances with large KK number tend to get smeared out into a continuum. 
Present searches for graviton resonances at the Tevatron as well as analyses 
of their 
indirect contributions to electroweak observables already place significant 
constraints on the $c-m_1$ plane. When combined with our theoretical 
prejudices the complete allowed region for the RS model is shown in Fig.~1 in 
comparison to the reach of the LHC. Even given some fuzziness in our 
prejudices it is apparent that the LHC should cover the entire RS parameter 
space either by discovering a graviton resonance or excluding the model. 
\begin{figure}
\centerline{
\psfig{figure=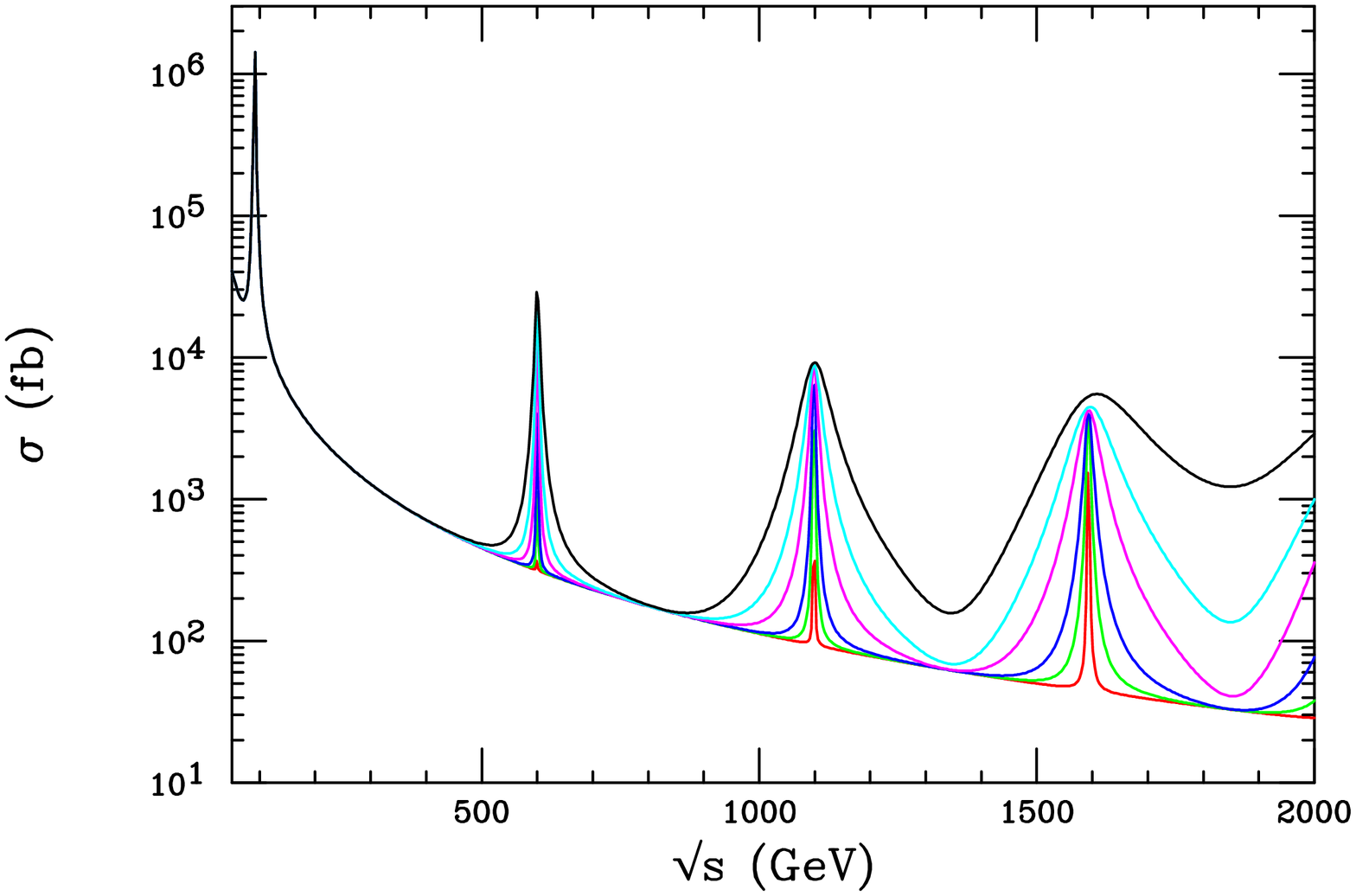,height=7.cm,width=7.0cm,angle=90}
\hspace*{5mm}
\psfig{figure=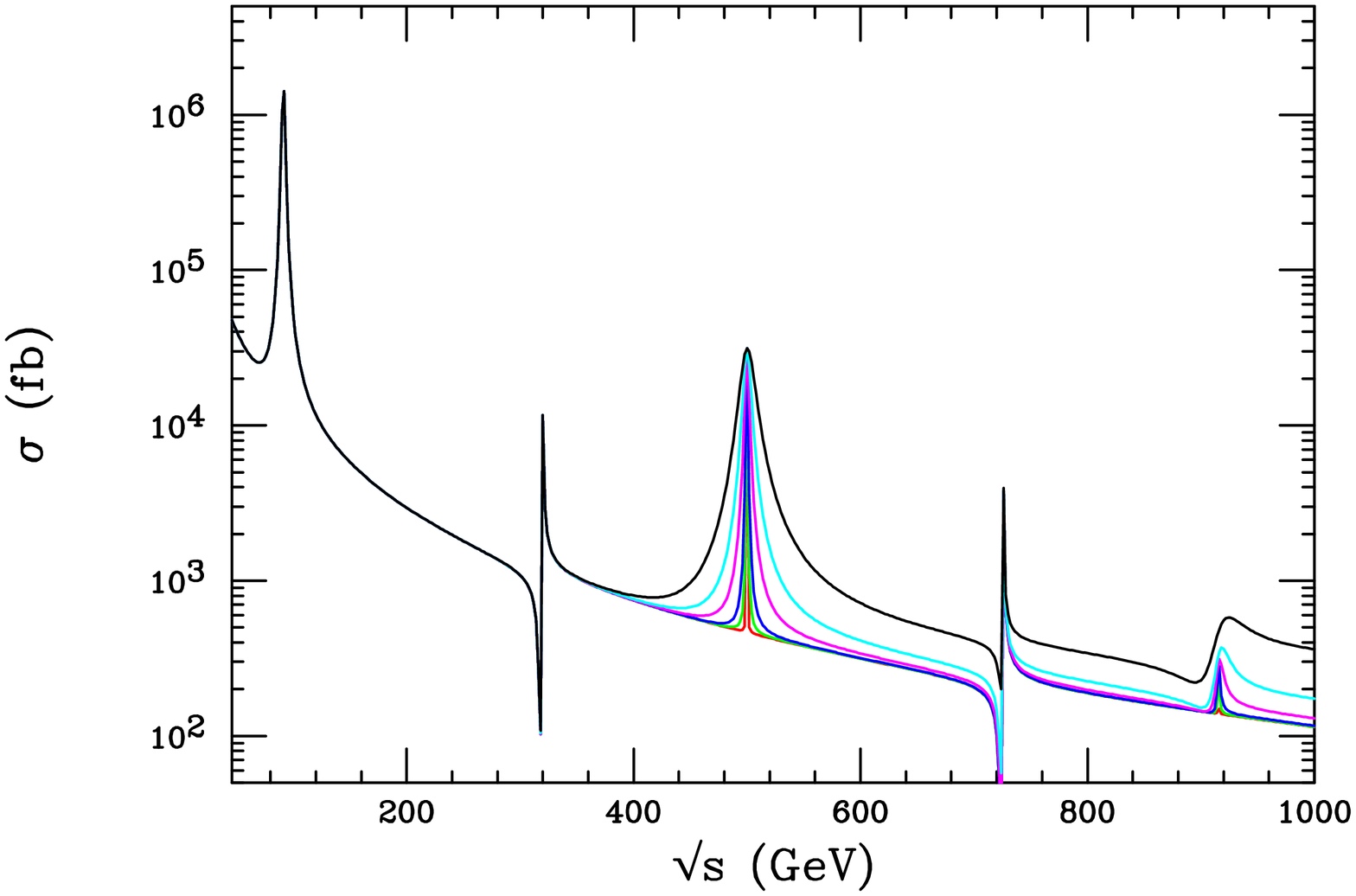,height=7.cm,width=7.0cm,angle=90}}
\vspace*{-0.05cm}
\caption{The left panel shows the production of KK graviton resonances in the 
process $e^+e^-\to \mu^+\mu^-$ assuming $m_1=600$ GeV for various values of 
$c$. In the right panel one sees the simultaneous 
production of graviton and gauge KK states typical of 
regions II and III via the 
process $e^+e^- \to \mu^+\mu^-$.}
\label{fig1}
\end{figure}
If the SM gauge fields alone are allowed to propagate in the bulk then it can 
be shown that the gauge KK excitations couple much more strongly to the 
remaining wall fields than do the zero modes{\cite {dhr}} by a factor 
$\simeq \sqrt {2\pi k r_c}$. The exchange and mixing of these modes contribute 
to the electroweak observables and result in a bound $\Lambda_\pi > 100$ TeV 
which is perhaps too high to claim a solution to the hierarchy problem. This 
strong bound can be alleviated by also placing the SM fermions in the bulk as 
well with the Higgs field remaining on the wall for a number of technical 
reasons{\cite {dhr}}. 
For simplicity and to avoid FCNC we assume that all SM fermions have an 
identical 5-d mass $m_{5d}=k\nu$, with $\nu$ a parameter of order unity. 
Specifying $\nu$ and $m_1$ for the graviton determines all of the KK masses 
with fermion excitations always more massive than gauge excitations and are 
approximately linear functions of $|\nu+1/2|$. 
For the phenomenology of Large Extra Dimensional (LED) physics \cite{ADD},
please look at \cite{ADD-phen} for single
graviton emission processes  as missing energy events and  
the indirect effects of the massive graviton exchange 
on various collider experiments.
\\

\paragraph*{Torsion Induced Interaction in LED}\cite{Tak}.
The general situation, in which the Lorentz group is gauged as well, will
give rise to an antisymmetric part to the connection coefficient, the torsion
tensor.  This general
situation obtains in the presence of intrinsic spin; the torsion tensor is then
coupled to the intrinsic spin current, which then represents yet another source
of gravity.  Since this current cannot be eliminated
by a choice of coordinates, the situation breaks the strong form of the
equivalence principle.  It is possible to require that the
torsion tensor still vanishes, but this demand will need to be preserved under
radiative corrections by invoking additional symmetries.  
Here the consequences of torsion in the
context of a model with large extra dimensions are considered.
Allowing the torsion tensor
$T^{\alpha}{}_{\beta\gamma} = 
\tilde{\Gamma}^{\alpha}{}_{\beta\gamma}-
\tilde{\Gamma}^{\alpha}{}_{\gamma\beta}$ to be non--zero introduces
an extra piece into the gravitational connection
\begin{equation}
\tilde{\Gamma}^{\alpha}{}_{\beta\gamma}=
\Gamma^{\alpha}{}_{\beta\gamma} + K^{\alpha}{}_{\beta\gamma},
\end{equation}
where $\Gamma^{\alpha}{}_{\beta\gamma}$ is the usual metric
contribution,
and $K_{\alpha\beta\gamma} = \frac{1}{2}\left( T_{\alpha\beta\gamma}
-T_{\beta\alpha\gamma}-T_{\gamma\alpha\beta} \right)$ is known
as the contorsion tensor.
The action of the model is given by
\begin{eqnarray}
\lefteqn{
S = -\frac{1}{\hat{\kappa}^2} \int d^{4+n}x \;
     \sqrt{\vert \hat{g}_{4+n}\vert} \;\tilde{R}
} & &  \label{action}
\\
& + & \int d^{4}x \;\sqrt{\vert \hat{g}_4 \vert}\; \frac{i}{2}
         \left[ \bar\Psi\gamma^\mu\tilde\nabla_\mu\Psi
              - \left( \tilde\nabla_\mu\bar\Psi \right)\gamma^\mu\Psi
              + 2 i M \bar\Psi \Psi
         \right].
\nonumber	
%\label{action}  
\end{eqnarray}	
Here $\hat{\kappa}^2=16\pi G_N^{(4+n)}$, $\tilde{R}$ is the $4+n$
dimensional scalar curvature, and $\hat{g}_{4+n}$ and $\hat{g}_4$
are respectively the $4+n$ and $4$--dimensional (induced) metric
determinants.  
Elimination of torsion from the action by imposing the equations
of motion results in \cite{Chang:2000yw}:
\begin{eqnarray}
S & = & -\frac{1}{\hat{\kappa}^2} \int d^{4+n}x
\;\sqrt{\vert \hat{g}_{4+n} \vert}\;R
        \\  \label{action2}
  &   & + \int d^{4}x \;\sqrt{\vert \hat{g}_4 \vert}\;
        \left[ \bar\Psi\left( i\gamma^{\mu}\nabla_{\mu} - M \right)\Psi
             + \frac{3}{32} 
               \frac{ \sqrt{ \vert \hat{g}_4 \vert } }{ \sqrt{\vert
\hat{g}_{4+n} \vert} } 
               \hat{\kappa}^2\,
               \left( \bar\Psi\gamma_{\mu}\gamma_5\Psi \right)^2\,
               \delta^{(n)}(0) 
        \right]\;. \nonumber
\end{eqnarray}
where $R$ is the torsion-free curvature.
The delta--function appearing in this expression should be regularized
to account for a finite wall thickness.
As a result, the
leading ${\cal O}\left( \hat{\kappa}^2 \right)$ torsion contribution
to the action is given by
\begin{equation}
\Delta S
= \int d^4x\;\frac{3\pi}{n M_S^2}
      \left[ \sum_j \bar\Psi_j\gamma_{\mu}\gamma_5\Psi_j
      \right]^2\;,
\label{action3}
\end{equation}
where $j$ runs over all fermions existing on the wall.
The expansion in $\hat{\kappa}$ is expected to be valid provided the
typical energy $E$
of a physical process is  below the cutoff scale $M_S$.
The torsion induced contact interaction Eq.~(\ref{action3}) can
be constrained through its effect on
$Z$--pole electroweak observables.
The corrections  shifts the $Z$--couplings. 
Performing a global fit to the LEP/SLD electroweak observables
will lead to a constraint on $Z$--couplings, which in turn will
give us a limit on $M_S$.
\\

\paragraph*{Extroweak Symmetry Breaking and LED}\cite{Che}.
The electroweak symmetry may be broken by a composite Higgs which 
arise naturally as a bound state of the top quark if the standard 
model gauge fields and fermions propagate in extra dimenions.  The top 
quark mass and the Higgs mass can be predicted from the infrared fixed 
points of the renormalization group equations. The top quark mass is 
in good agreement with the  experimental value, and the Higgs boson 
mass is predicted to be $\sim 200$ GeV \cite{CDH}. The bounds on the 
compactification scale can be quite low if all standard model fields 
propagate in the same extra dimensions due to the momentum 
conservation in extra dimensions. The current lower  limits are about 
300 GeV for one extra dimensions and 400-800 GeV for  two extra 
dimensions. The future collider experiments may either discover the 
Kaluza-Klein (KK) states of the standard model fields or raise their 
mass limits significantly.  There may also be some other light bound 
states which could be observed at upcoming collider experiments. 
Compared with the usual four-dimensional dynamical electroweak symmetry 
breaking (EWSB) 
models, the higher-dimensional model has the advantage that 
the binding force can be the SM gauge interactions themselves, 
without the need of introducing new strong interactions. In addition, 
it also gives a prediction 
of the top quark mass naturally in the right range. In the minimal 
four-dimensional top quark condensate model, the top quark is too 
heavy, $\sim 600$ GeV,  
if the compositeness scale is in the TeV range. 
With extra dimensions, the KK excitations of the top quark also 
participate in the EWSB, so the top quark mass can be smaller. 
Another way of understanding of the top Yukawa coupling being $\sim 1$ 
instead of the strong coupling value $\sim 4\pi$ is that (the zero 
mode of) the top quark coupling receives a volume dilution factor 
because it propagates in extra dimensions.  In fact, the top quark 
mass can be predicted quite insensitively to the cutoff because of the 
infrared fixed point behavior of the  renormalization group (RG) 
evaluation. The infrared fixed point is rapidly approached due to the 
power-law running in extra-dimensional theories. even 
though the cutoff scale is not much higher the the weak 
scale. Similarly, the Higgs self-coupling also receives  the 
extra-dimensional volume suppression. As a result, the physical Higgs 
boson is relatively light, $\sim$ 200 GeV, in contrast with the  usual 
strongly coupled four-dimensional models. It is also governed by the 
infrared fixed point of the RG equations. 
 
\section*{IV. Comments on Gravity}

As summarized in previous sections, the main theme of our Working Group
was ``How precisely has Gravity been probed microscopically and
macroscopically?"
Here I note about it from two different approaches.
\\

\paragraph*{CPT violation from a tilted brane}\cite{CTP}.
The tilted brane in large extra dimension can be described by a
low energy effective theory. In this theory, 
the graviphoton ${\cal A}_{\mu}(x)$ obtains a mass
$m_{\cal A}\!\sim\!1 \ {\rm mm}^{-1}$ by eating up the 
corresponding Goldstone mode. After integrating
out the graviphoton, the effective theory for the other 
pseudo-Nambu-Goldstone boson(s)  $\chi$ 
describing the dynamics of the brane
is represented by the four-dimensional Lagrangian density
\begin{equation}
{\cal L}_{\rm brane} = 
g^{\mu \nu} {\partial}_{\mu} {\chi}  {\partial}_{\nu} {\chi}  \ ,
\label{Lbrane}
\end{equation}
where $g_{\mu \nu}$ is the induced metric on the brane.
The tilted brane solution to the equation of motion
${\partial}^2 {\chi}\!=\!0$ is
\begin{equation}
\chi = \sqrt{T} {\alpha} x^j \ ,
\label{tbsol}
\end{equation}
where ${\hat x}^j$ is the direction along which the tilting occurs,
$T$ is the brane tension, i.e., the energy per
3--space unit volume, and $\alpha$ is the angle of the
tilting. The tiny tilting angle $\alpha$ causes the brane to be
``stretched'' by the factor of $1\!+\!{\alpha}^2/2$.
On the other hand,
${\alpha}\!=\!T^{-1/2} {\partial} {\chi}/{\partial} x^j$ 
by (\ref{tbsol}). This implies that the induced metric
$g_{\mu \nu}$ and the flat (untilted) metric
$g^{(0)}_{\mu \nu}$ are related
\begin{equation}
g_{\mu \nu} = g^{(0)}_{\mu \nu} + \frac{1}{2} T^{-1}
{\partial}_{\mu} {\chi} {\partial}_{\nu} {\chi} \ .
\label{indmetr}
\end{equation}
An analogous relation for the basis
4--vectors $e_a$ on the tilted brane
\begin{equation}
e_a^{\mu} = g^{(0) \mu}_a + \frac{1}{4} T^{-1}
{\partial}^{\mu} {\chi} {\partial}_{a} {\chi}
\label{eaind}
\end{equation}
follows from (\ref{indmetr}) due to $e_a \cdot e_b = g_{a b}$.
The kinetic term for the fermionic fields on the
tilted brane involves 
${\partial \llap /}_{\rm tilted} = e_a^{\mu} {\gamma}^a {\partial}_{\mu}$.
Therefore, when using (\ref{eaind}), the kinetic energy term in
the tilted brane background can be rewritten as
\begin{eqnarray}
{\cal L}_{\rm kin.} & = & 
({\overline \psi} {\partial \llap /} {\psi})_{\rm tilted}
= {\overline \psi} {\gamma}^{\mu} {\partial}_{\mu} {\psi}
+ \frac{1}{4} T^{-1}  \left( 
{\partial}^{\mu} {\chi} {\partial}_{\nu} {\chi} \right)
\left( {\overline \psi} {\gamma}^{\nu} {\partial}_{\mu} {\psi} \right) \ ,
\label{induced}
\end{eqnarray}
where all the derivatives are in the flat metric. 
If we now expand around the tilted brane solution (\ref{tbsol}), i.e., 
$\chi\!=\!\sqrt{T} {\alpha} x^j\!+\!\delta \chi$,
we obtain from the last term of (\ref{induced}) 
interaction terms which break the Lorentz and rotational
invariance
\begin{eqnarray}
\delta {\cal L} &=& \frac{1}{4} T^{-1/2} {\alpha}
{\partial}_{\nu} ( \delta \chi ) \left[ {\overline \psi}
\left( {\gamma}^{\nu} {\partial}^j\!+\!{\gamma}^j {\partial}^{\nu}
\right) \psi \right] + \frac{1}{4} {\alpha}^2 
{\overline \psi} {\gamma}^{\nu} {\partial}^{\mu} \psi 
{\big |}_{\nu=\mu=j} \ .
\label{liviol}
\end{eqnarray}
The first term ($\propto\!{\alpha}$), in addition, violates
CPT, because ${\partial}_{\nu} ( \delta \chi )$ is
odd and the term in $[ \ldots ]$ is even under CPT.
\\

\paragraph*{Neutrino oscillation from Violation of Equivalence Principle
}\cite{Leung}. If $\gamma_\nu \equiv \sqrt{1-\beta_\nu^2}$ is flavor
dependent, then different neutrinos will undergo different gravitational time
delays when passing through the same gravitational potential and thereby
acquire different phase shifts.
These phase shifts are observable
owing to the difference in  the  particle  bases  that
diagonalize the weak  and  the  gravitational  interactions. 
As a consequence, a $\nu_e$
will be able to oscillate into $\nu_\mu$.  
In the absence of non-gravitational interactions, the properties of a
spin-1/2 particle in a specified gravitational field, $G_{\alpha\beta}$,
are usually described to first order (linearized therory) by the interaction
Lagrangian density
\begin{equation}
{\cal L}_{{\rm int}} = i \frac{f}{4} G^{\alpha\beta} [\bar{\psi} \gamma_\alpha
 \partial_\beta \psi - (\partial_\alpha \bar{\psi}) \gamma_\beta \psi] ,
\label{lag}
\end{equation}
where $f = \sqrt{8 \pi G_N}, G_N$ is Newton's constant and the metric
of flat space is $g_{\alpha\beta} = (+1,-1,-1,-1)$.  
An interaction of the above form but one which allows the neutrinos
$\nu_1$ and $\nu_2$ to couple to gravity with different strengths $f_1$ and
$f_2$ can be postulated. Then,
the postulated interaction leads to the equations of motion for the
{\it massless} neutrino fields, $\nu_j$,
\begin{equation}
[(g^{\alpha\beta} + \frac{f_j}{2} G^{\alpha\beta}) \gamma_\alpha
\partial_\beta + \frac{f_j}{4} (\partial_\alpha G^{\alpha\beta})
\gamma_\beta] \nu_j = 0, ~~~~~j = 1, 2, ...
\label{EoM}
\end{equation}
In this case, the $\nu_j$ satisfy a Klein-Gordon equation,
\begin{equation}
(g^{\alpha\beta} + f_j G^{\alpha\beta}) \partial_\alpha \partial_\beta
\nu_j = 0 .
\label{KG}
\end{equation}
If we assume the gravitational field is determined by a static
macroscopic matter distribution in the harmonic gauge,   such a
field is given in terms of the Newtonian potential $\phi$ by
\begin{equation}
G_{\alpha\beta} = 2 \phi \delta_{\alpha\beta} / f  ,
\label{hgauge}
\end{equation}
where $\phi(\infty) \rightarrow 0$.
To illustrate the essential properties of the resulting phase shifts,
we consider the case of constant $\phi$, where we have
the energy-momentum relation
\begin{equation}
E^2 (1 + 2 \gamma_j \phi) = p^2 (1 - 2 \gamma_j \phi) .
\label{E2}
\end{equation}
For the simple case of two neutrinos, this implies that, after traversing
a distance $l$, the two components,
($\nu_1, \nu_2$), of a state $\nu_e$ will develop a phase difference of
$\delta = 2 (\gamma_1 - \gamma_2) \phi l p$.  
If we compare this phase shift with that obtained in the well known
case of vacuum oscillations induced by a neutrino mass difference,
we find that they are related by the formal connection,
\begin{equation}
\frac{\Delta m^2}{2E} \rightarrow 2E |\phi| \Delta\gamma ,
\label{sub}
\end{equation}
where $\Delta\gamma \equiv \gamma_2 - \gamma_1$.  By analogy,
the $\nu_e$ survival probability after traversing a distance
$l$ is given by
\begin{equation}
P(\nu_e \rightarrow \nu_e) = 1 - \sin^2 2\theta_G \sin^2 \frac
{\pi l}{\lambda}, ~~{\rm with}~~
\lambda = 6.2~{\rm km} \Bigl(\frac{10^{-20}}{|\phi \Delta\gamma |}\Bigr)
\Bigl(\frac{10~{\rm GeV}}{E}\Bigr) .
\label{lambda}
\end{equation}
\\

\paragraph*{Acknowledgements.~~} 
I would like to thank the co-convenors of P5-WG, S. Tkaczyk and G. Wilson.
The work was supported in part by Seo-Am foundation,
in part by  BK21 Program, SRC Program and Grant No. 2000-1-11100-003-1
of the KOSEF, and in part by the KRF Grants, Project No. 2000-015-DP0077.

\def\MPL #1 #2 #3 {Mod. Phys. Lett. {\bf#1},\ #2 (#3)}
\def\NPB #1 #2 #3 {Nucl. Phys. {\bf#1},\ #2 (#3)}
\def\PLB #1 #2 #3 {Phys. Lett. {\bf#1},\ #2 (#3)}
\def\PR #1 #2 #3 {Phys. Rep. {\bf#1},\ #2 (#3)}
\def\PRD #1 #2 #3 {Phys. Rev. {\bf#1},\ #2 (#3)}
\def\PRL #1 #2 #3 {Phys. Rev. Lett. {\bf#1},\ #2 (#3)}
\def\RMP #1 #2 #3 {Rev. Mod. Phys. {\bf#1},\ #2 (#3)}
\def\NIM #1 #2 #3 {Nuc. Inst. Meth. {\bf#1},\ #2 (#3)}
\def\ZPC #1 #2 #3 {Z. Phys. {\bf#1},\ #2 (#3)}
\def\EJPC #1 #2 #3 {E. Phys. J. {\bf#1},\ #2 (#3)}
\def\IJMP #1 #2 #3 {Int. J. Mod. Phys. {\bf#1},\ #2 (#3)}


\begin{references} 
 
\bibitem{Dom} D. Dominici, Proc. of LCWS2000@FNAL, Oct. 2000.    
\bibitem{Kal} P. Kalyniak, Proc. of LCWS2000@FNAL, Oct. 2000.
\bibitem{Rie} S. Riemann, Proc. of LCWS2000@FNAL, Oct. 2000.
\bibitem{Bat} M. Battaglia, Proc. of LCWS2000@FNAL, Oct. 2000.
\bibitem{Ily} V.A. Ilyin, Proc. of LCWS2000@FNAL, Oct. 2000.
\bibitem{Han} T. Han, Proc. of LCWS2000@FNAL, Oct. 2000.
\bibitem{Heu} C. Heusch, Proc. of LCWS2000@FNAL, Oct. 2000.
\bibitem{Lyk} J. Lykken, Proc. of LCWS2000@FNAL, Oct. 2000.
\bibitem{Hew} J. Heweet, Proc. of LCWS2000@FNAL, Oct. 2000.
\bibitem{Dav} H. Davoudiasl, Proc. of LCWS2000@FNAL, Oct. 2000.
\bibitem{Riz} T. Rizzo, Proc. of LCWS2000@FNAL, Oct. 2000.
\bibitem{Tak} T. Takeuchi, Proc. of LCWS2000@FNAL, Oct. 2000.
\bibitem{Che} H.C. Cheng, Proc. of LCWS2000@FNAL, Oct. 2000.
\bibitem{dbess}  R.~Casalbuoni {\it et al.}, \PLB 349 533 1995. 
\bibitem{wp1} 
S. Godfrey, P. Kalyniak, B. Kamal, and A. Leike, \PRD 61 113009 2000 . 
\bibitem{wp2} 
S. Godfrey, P. Kalyniak, B. Kamal, M. Doncheski and A. Leike, hep-ph/0008157. 
\bibitem{3fm} 
S. Chivukula, E. Simmons and J. Terning, \PLB 331 383 1994 . 
\bibitem{Noj} M. Drees and M.M. Nojiri, \PRD 49, 4595 1994 .
\bibitem{heuschmink} C.A. Heusch, P. Minkowski,  \PLB 374 116 1996 .
\bibitem{emu} G. Cvetic, C.S. Kim and C.W. Kim, \PRL 82 4761 1999 ;
G. Cvetic and C.S. Kim, \PLB 461 248 1999 .
\bibitem{RS} L. Randall and R. Sundrum, \PRL 83 3370 1999 .
\bibitem{dhr} H. Davoudiasl, J.L. Hewett and T.G. Rizzo, \PRL 84 2080 2000 ;
              \PLB B473 43 2000 .
\bibitem{ADD} N. Arkani-Hamed, S. Dimopoulos and G. Dvali, \PLB 429 263 1998 .
\bibitem{ADD-phen} E.A. Mirabelli, M. Perelstein and M.E. Peskin, 
\PRL 82 2236 1999 ; J. L. Hewett, \PRL 82 4765 1999 ; 
C. Balazs, H.-J. He, W.W. Repko and C.-P. Yuan, \PRL 83 2112 1999 ; 
K. Cheung and W.-Y. Keung, \PRD 60 112003 1999 ; 
K. Agashe and N. G. Deshpande, \PLB B456 60 1999 ; 
T. G. Rizzo and J. D. Wells, \PRD 61 016007 2000 ; 
K. Y. Lee, H. S. Song and J. Song, \PLB 464 82 1999 ; 
C.S. Kim, K.Y. Lee and J.H. Song, hep-ph/0009231.
\bibitem{Chang:2000yw}
L.N. Chang, O. Lebedev, W. Loinaz and T. Takeuchi, \PRL 85 3765 2000 .
\bibitem{CDH} 
H.-C.~Cheng, B.~A.~Dobrescu and C.~T.~Hill, \NPB 589 249 2000 . 
\bibitem{CTP} G. Cvetic, S.K. Kang, C.S. Kim and K. Lee, \PRD 62 057901 2000 .
\bibitem{Leung} A. Halprin, C.N. Leung and J. Pantaleone, \PRD 53 5365 1996 .

\end{references}
\end{document}